\documentclass[aps,twocolumn]{revtex4}
\usepackage{amsmath}
\usepackage{amssymb}
\usepackage{graphicx}
\usepackage{floatrow}

\begin{document}

\title{Active Learning for Rapid Targeted Synthesis of Compositionally Complex Alloys}

\author{Nathan S. Johnson}
\email{natej@stanford.edu}
\affiliation{SLAC National Accelerator Laboratory, Menlo Park, CA 94025} 

\author{Aashwin Ananda Mishra}
\affiliation{SLAC National Accelerator Laboratory, Menlo Park, CA 94025} 

\author{Apurva Mehta}
\affiliation{SLAC National Accelerator Laboratory, Menlo Park, CA 94025}

\begin{abstract}
The next generation of advanced materials is tending toward increasingly complex compositions. Synthesizing precise composition is time-consuming and becomes exponentially demanding with increasing compositional complexity. An experienced human operator does significantly better than a beginner but still struggles to consistently achieve precision when synthesis parameters are coupled. The time to optimize synthesis becomes a barrier to exploring scientifically and technologically exciting compositionally complex materials. This investigation demonstrates an Active Learning (AL) approach for optimizing physical vapor deposition synthesis of thin-film alloys with up to five principal elements. We compared AL based on Gaussian Process (GP) and Random Forest (RF) models. The best performing models were able to discover synthesis parameters for a target quinary alloy in 14 iterations. We also demonstrate the capability of these models to be used in transfer learning tasks. RF and GP models trained on lower dimensional systems (i.e. ternary, quarternary) show an immediate improvement in prediction accuracy compared to models trained only on quinary samples. Furthermore, samples that only share a few elements in common with the target composition can be used for model pre-training. We believe that such AL approaches can be widely adapted to significantly accelerate the exploration of compositionally complex materials. 

\end{abstract}
\date{\today}
\maketitle

\section{Introduction}

Traditional alloy engineering mixes small additions of alloying elements into a primary element matrix for performance improvement.  Still, after centuries of incremental improvements, we are rapidly reaching the limit of performance from primary alloys.  Over the last decade, compositionally complex alloys, sometimes called multi-principal element alloys or high entropy alloys [7–10], containing many (3+) elements in significant proportions, have shown outstanding properties for a wide range of engineering applications, including structural alloys \cite{Kim2019, Rar2004}, batteries \cite{Wang2022}, thermoelectrics \cite{Anand2018}, shape-memory alloys \cite{Zarnetta2010, AlHasan2020}, catalysts \cite{Li2022}, high entropy alloys \cite{Jung2022,Yan2018, Zhou2022, Liang2023, Han2023}, high entropy ceramics \cite{Oses2020} and more. Many of the desired alloys are composed of refractory and low-melting elements, and the final composition is seldom the same as the composition of the input reactant; it takes several iterations before the desired composition is reached.  Discovering and fabricating precise alloy compositions in these high-dimensional spaces using a traditional approach is substantially slower and more expensive than desired.  

The deposition of one element, and consequently reaching the desired alloy composition, is often influenced by the deposition of the other elements; therefore, a higher compositional complexity often means a significantly more complex synthesis optimization in coupled high-dimensional parameter space.  The problem is further exacerbated because these new functional alloys are needed as thin films for catalysts and coatings or desired to be fabricated by advanced manufacturing methods such as additive manufacturing or electroplating.  High-throughput synthesis and characterization, guided by physical models, machine learning, or intuition (and human expertise), are suggested as a path for accelerated search of complex systems \cite{Gregoire2023}. 
 
Currently, synthesis conditions are arrived at by a human operator relying on expertise (or intuition) in assessing the coupling between different elemental dimensions through trial iterations.  An expert operator usually finds synthesis parameters for binary or ternary systems to a few percent of the desired compositions in a few ($< 5$) trial iterations.   However, a human operator's biggest challenge is learning the complex coupling as the dimensionality increases.  They struggle to improve the precision beyond a few percent and require exhaustive calibrations and iterative parameter tuning, especially if the coupling between elements is complex (non-linear).  A common strategy employed to combat the curse of dimensionality is to reduce the problem's dimensionality and then add additional dimensions one at a time.  For example, instead of synthesizing a 5-element sample immediately, researchers might first manufacture 3- and 4-element sub-alloys.  This multi-step approach allows researchers to tune the composition by just a few elements at a time instead of trying to tune five elements simultaneously.  This approach converges if the additional dimensions are weakly coupled, and the challenge often is to find a strongly coupled base subset of the target space and separate it from the weakly coupled one if it exists. 

Another common approach is physics- or chemistry-based models to map the complex coupling between deposition parameters.  Bunn et al. demonstrated a computationally fast continuum model for optimizing film thickness in thin film samples synthesized via magnetron sputtering \cite{Bunn2016}.  Their method also requires very few initial data points before achieving high prediction accuracy.  Furthermore, their approach demonstrates high interpretability, directly reporting parameters like gun power and angle.  However, Bunn's model was shown for thickness measurements only, has yet to be applied to composition optimization, and does not incorporate multiple elements.  The physiochemical modeling approach is compelling, and in many ways it quantifies information that a human acquires to build intuition. However, physiochemical models work when there is a substantial theoretical understanding of the synthesis process.  Often, however, deep theoretical understanding is not available; what is available is empirical observations from trial synthesis.  For example, Xia et al. detail latent causes that alter sputtering rates in magneto sputtering of multi-element thin films and qualitative insights on how the composition changes when multiple sources are used simultaneously \cite{Xia2020}.  However, quantitative prediction from theory of different sputtering rates needed in multi-elemental synthesis to reach the desired chemistry is very challenging.

Efforts to incorporate the empirical information from human operators to formulate quantitatively accurate models for higher dimensional target spaces are important. There are several traditional empirical methods for process optimization.  One such method for magnetron sputtering is well-detailed in an article by Alami et al. \cite{Alami2006}, which involves depositing at power/sputtering rate steps and measuring the composition at each step.  This exhaustive empirical method works well for low-dimensional systems like binary alloys.  However, as the number of elements grew or a finer composition control was needed, the number of empirical measurements became burdensome.  Another category of process optimization of sputtered thin films measures individual elements' sputtering rates at various powers and angles.  A sensor, like a quartz crystal monitor (QCM), can measure the sputtering rate directly as a function of cathode power and gun angle \cite{Deki1997}.  The sputtering rate of each element can be set to achieve the desired atomic percent of that element in the final film.  Measuring individual sputtering rates on a QCM requires fewer measurements than exhaustively going through all power and angle combinations for an n-element system.  However, accurate QCM measurements require knowledge of the sputtering elements' Z-number and the deposited film's density.  The sputtering rates of an $n$-ary system are tricky to measure using a quartz crystal monitor because the density of the alloyed system is a) usually not known a priori and b) changes as a function of the sputtering rate of each element.  Furthermore, the number of sputtering rates and interaction terms grows as $\left(n^2 + n\right)/2$ for $n$ many elements.  QCM measurements also do not capture the interactions occurring when multiple sputtering sources are turned on simultaneously. The traditional empirical methods are data-hungry and, therefore, not very useful when exploring a new compositional chemistry.

Machine learning (ML) approaches such as Active learning (AL) and Transfer learning (TL), provide tools that allow empirical methods to start from the earliest stages of exploration when information and insights about a newly discovered target space are minimal \cite{Gongora2020,Xue2016, Kusne2020,Nikolaev2016, Ament2023}.  These approaches overcome many challenges a human operator faces, including optimizing in high-dimensional target spaces and the ability to transfer knowledge gained from one system to another quickly.  As the exploration progresses, it also provides real-time insights into the structure of the target space, including the strength of coupling between dimensions and identification of a lower dimensional strongly coupled target subspace, insights that the operator can exploit to fine-tune the exploration strategy further.

In this article, we illustrate these approaches for exploring and optimizing magnetron-sputtered synthesis of 5-element alloys containing refractory and volatile elements.  Although our results are specific to a particular synthesis tool and limited target composition spaces, the insights that have emerged from these studies and the approaches developed here are widely applicable.  We will discuss the insights as they emerge and highlight how they can be broadly applied to transform research in compositionally complex alloys in the concluding section of this work.

\begin{figure*}[t]
    \centering
    \includegraphics[width=1\linewidth]{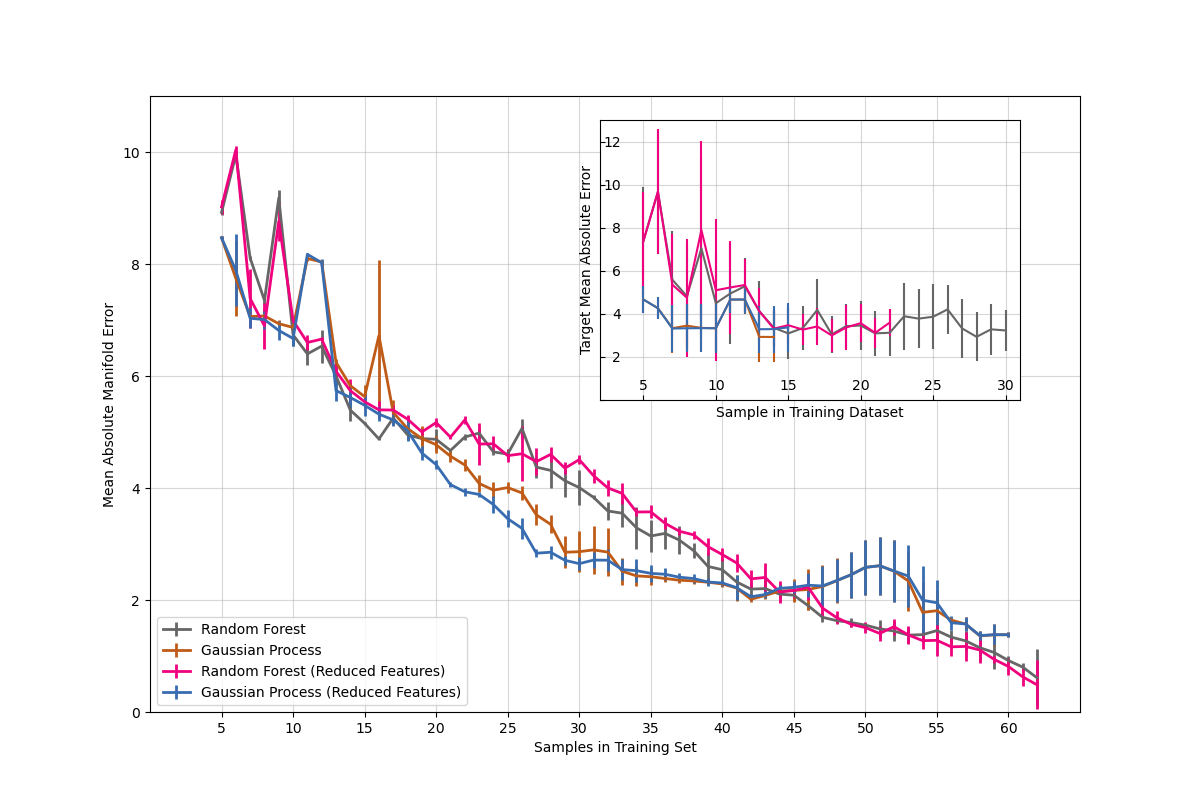} 
    \caption{Mean absolute error for a quinary composition as a function of the number of training samples. (inset) Mean absolute error for a given target composition as a function of the number of training samples. Error bars represent the standard deviation in prediction error across ten different target composition predictions.} 
    \label{fig:all_models_total_errors}
\end{figure*}

\section{Results and Discussion}

\begin{figure*}
    \includegraphics[width=1\linewidth]{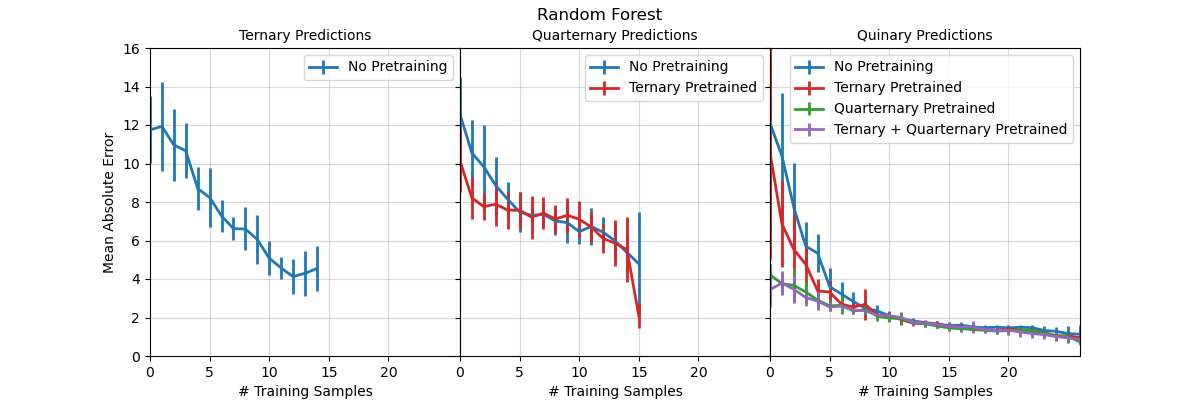}
    \caption{Performance of active learning models trained on successively more complex compositions. Plots are organized based on the composition being predicted; the leftmost plot are predictions for quinary compositions; quarternary in the middle; ternary on the right.}
    \label{fig:transfer_learn}
\end{figure*}

Active learning was implemented through two regression models: a Gaussian process regression (GPR) and random forest (RF). These specific models were selected based on a set of rationale. Random Forests and Gaussian Process based regression models have been observed to be very effective at modeling with tabular data in prior research \cite{grinsztajn2022tree, caruana2006empirical}. Additionally, these are very popular for Active Learning applications, for instance, Gaussian Process models are the default surrogate model in Bayesian Optimization studies \cite{rasmussen2006gaussian}. The models were trained on a dataset of sputtering synthesis parameters and compositions for ternary, quarternary, and quinary alloys. At each learning iteration, the models were queried for the next datapoint to test based on a maximum uncertainty (MU) schema. The goal was to correctly predict the synthesis parameters for a target alloy using as few training datapoints as possible.

Model performance was assessed two ways: the models' ability to correctly predict a target composition (inset of Figure \ref{fig:all_models_total_errors}) and its error in predicting all compositions across the composition manifold (main graph on Figure \ref{fig:all_models_total_errors}). Active learning models were able to find a target quinary composition after only 14 sample iterations. For predicting target compositions the models were terminated after the Mean Absolute Error (MAE) reached 3\% since this is the uncertainty level of the composition measurement. The best performing model was able to correctly predict synthesis parameters across the entire composition manifold to within 3\% error after only 26 iterations.

The error is calculated as the absolute difference in the target atomic percentage $\mathbf{Y}$ and the measured atomic percentage $\hat{\mathbf{Y}}$. This error is $|Y_i-\hat{Y_i}|$ for the $i$th element in an $n$-ary composition. For a complete sample, the error is the mean of all differences $ (1/n) \sum_{i}^{n} |Y_i - \hat{Y_i}|$ for all elements, often referred to as the Mean Absolute Error (MAE). In this study the training samples were ternary, quarternary, or quinary transition metal alloys. The target compositions in Figure \ref{fig:all_models_total_errors} were all quinary alloys selected from 6 possible elements: Titanium, Vanadium, Niobium, Tantalum, Antimony, and Iron. The manifold error is taken over all compositional complexity; the main graph in Figure \ref{fig:all_models_total_errors} represents the error in prediction for ternary, quaternary, and quinary alloys. The error shown is the average of 10 model runs with each run sampling a different initial datasets. The shown error bars are the 1st standard deviation of error across all 10 runs.

The models' prediction ability stands in contrast to an expert human operator performance; experts typically require 20 or more iterations to synthesize one target quinary alloy correctly. However, a human expert learns synthesis parameters for one alloy and can predict parameters for alloys with incrementally different chemistry but struggles to predict a significantly different composition in the same quinary composition space. In contrast, the models learn the full composition space.

The sparsity of training data was not a barrier to high predictive performance in either RF or GPR models. Magnetron sputtering synthesis, the method used in this study, is a labor intensive and slow process compared to others like spin coating; typically it takes a researcher up to two days to manufacture and characterize five samples. With these labor intensive manufacturing processes sparsity in data is a given. Researchers using active learning methods can still benefit from model-driven parameter guidance even with small dataset sizes.

As a baseline for data driven models, all of the AL based models were compared against a least squares based linear regression model. As shown in the supplemental information, the least squares regression model had absolute errors upwards of 15 after 5 training samples; this eventually decreased to an error of around 7 after all training samples had been fit to the model.

Models were also trained using a subset of the most important synthesis parameters, determined using the Mutual Information Index (MII) \cite{kraskov2004estimating} to test the impact of poorly-informative inputs on model performance. (MII is discussed in more detail below.) These models are labeled as `Reduced Features' in Figure \ref{fig:all_models_total_errors}. The performance of models with these reduced feature sets is within the bound of models trained on all features. This knowledge provides several practical insights for both the active learning model and the experiments. From a computational side, removing less informative features in studies with large feature sets can significantly reduce the computational cost of running an active learning workflow. From an experimental side, identification of less informative features helps reduce the complexity of synthesis studies. If a synthesis parameter is not informative of the desired measurement then it is better to fix that parameter and \textit{not} vary it at all. Additionally, the information regarding relative importances of input features for the model enables the domain scientist to compare the model's learned mapping to their understanding of the system\cite{arrieta2020explainable, adhikari2019leafage}. This allows the scientist to interpret the model's mapping and verify its rationale, leading to a higher degree of trust in the model\cite{davis2020measure}.


\subsection{Transfer Learning into Higher Dimensional Systems}
The most powerful feature of either RF or GPR model is their ability to learn synthesis conditions from previously made samples even if those samples are in a lower-dimensional composition space (ternary, quarternary), or share only a few elements in common with the target composition. Figure \ref{fig:transfer_learn} shows the MAE for RF models trained on successively more complex samples. Each MAE in Figure \ref{fig:transfer_learn} is assessed over all compositions at a given complexity (ternary, quarternary, quinary). 

Models trained only on ternary samples generally performed poorly, achieving a final MAE of $>3\%$ for predicting ternary compositions. Models for predicting quarternary samples trained on ternary samples show an immediate improvement in MAE over models trained only on quarternary samples. The biggest improvement is in quinary models trained on ternary, quarternary, or both systems. GPR models for predicting quinary compositions that were trained on ternary samples have an initial MAE of less than 10\%; the quinary prediction models trained on quarternary only or on both sub-systems showed an initial MAE less than 5\% and quickly approached an MAE of less than 2\%. The advantage of pre-taining, perhaps not surprisingly, is greatest at earlier stages (and sparser-data training stage) of learning. Pre-training on the lower dimensional spaces reduces the training time by nearly a factor of 2 for both quarternary and quinary composition spaces.  As synthesis of training/trial samples is slow and expensive, this is a significant saving, even though un-pre-trained models eventually achieve comparable accuracy.  The difference between the quinary models pre-trained on only quarternary and ternary+quarternary is marginal, suggesting that the quarternary model captures all of the significant relationships learned by the ternary model.

The GPR models showed similar performance to the RF models. A complementary plot to Figure \ref{fig:transfer_learn} for GPR models is included in the Supplemental Information.

In both cases, the model performance indicates that training on lower-dimensional samples is beneficial for predicting on high dimensional systems. Synthesis laboratories often have prior training data on lower dimensional systems already acquired. This prior data can seed AL based regression models. Furthermore, humans often work their way up to complex sample synthesis. Instead of trying to synthesize a 5-element sample immediately, researchers might first manufacture 3- and 4- element analogues. This allows researchers to tune the composition by just a few elements simultaneously instead of trying to tune 5 elements simultaneously. The active learning regression approach is compatible with this type of human calibration; as the human makes successively more complex samples, the regression model can be trained simultaneously. Once the human is prepared to make the most complex samples, they can immediately rely on the model predictions.  

\subsection{Feature Importance and Interdependence}
\begin{figure}
    \includegraphics[width=1\linewidth]{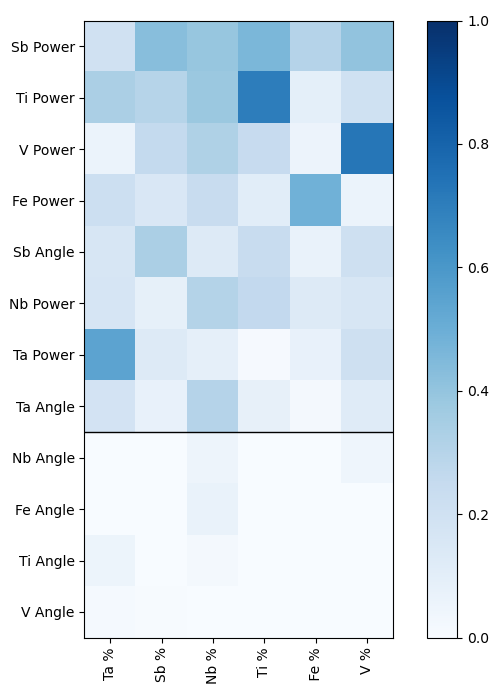}
    \caption{Mutual information index for all input parameters and atomic percentages for all data.}
    \label{fig:mutual_information}
\end{figure}

Going to higher composition spaces is significantly harder, not only because every additional element brings in additional parameters, but these parameters are often strongly coupled with the parameters from the lower dimensions.  Adding a new element to the composition spaces requires learning additional parameters and learning new interdependencies between already trained lower dimensional parameters with the new parameters. The interdependence of the synthesis parameters is best summarized in the mutual information index shown in Figure \ref{fig:mutual_information}. The mutual information index encodes how the knowledge of one variable decreases uncertainty about another variable. Unlike other correlation coefficients, it does not assume a linear relationship between the two variables.  

The mutual information index (MI) for a 12-dimension target space (gun power and gun angle for six elements) shows that 8 of those dimensions are strongly correlated.  The sputtering power for the elements is strongly correlated and strongly affects the atomic percentage of all the other elements. It is unsurprising that each element's gun power has the highest mutual information with its atomic percentage. Still, the MI with its atomic percentage is not uniformly high for every element.  For example, Sb gun power affects Ti atomic percentage as much as it does its own. However, the angles of the sputtering guns relative to the substrate play a negligible role for almost all elements. Only the angle of the Ta and Sb guns impacted the atomic percentage of other elements. This is likely due to the high Sb and Ta sputtering rate relative to all the other elements. The high sputtering rate of Sb and Ta means they can easily dominate the sample composition if they are both pointed directly at the substrate. Moving the angle of the gun away from the target is an effective way to modulate these high sputtering rates. For the other elements, whose sputtering rates are significantly lower than Sb or Ta, if they are pointed away from the substrate, their sputtering rate (and thus atomic percentage) quickly approaches zero. The MII for these values is shown below the black bar in Figure \ref{fig:mutual_information}. Their angle must be set so they are always pointed directly at the substrate. Regarding mutual information, this means the angle of Ta and Sb has a high MI with the final composition and all other elements have a low MII. 

The AL models most often suggested the same sputtering angle for the slowly sputtering V, Fe, Ti, and Nb (i.e., pointed directly at the substrate) even though there were training samples with other angles in the datasets. The model finds effective means of lowering the complexity of the target space without expert insight. It finds that when sputtering elements that have very different sputtering rates, it is often better to point the angle of slowly sputtering elements directly towards the substrate during a synthesis study and only adjust the angle of the high sputtering rate elements. The AL model, in effect, ``discovers" a good rule-of-thumb for human operators to follow when navigating a multi-element target space requiring very different sputtering rates. 

\subsection{Comparison to Other Active Learning Optimization Approaches}

The method presented herein shows high interpretability, high prediction accuracy, and a very low barrier to entry. The entire model can be executed in under 50 lines of code. The methods and classes used are well-documented online, with plenty of supporting tutorials. An example notebook and the full dataset are also included in this publication's supplemental material. There are also numerous open-source libraries and resources for performing the same process optimization outside of the ones used in this article.

The ability to perform transfer learning enables researchers to pre-train AL regression models even with data not directly relevant to the composition of interest. Exhaustive calibration studies, or QCM-guided calibration studies, can only make predictions \textit{within} the target system being calibrated. The AL regression approach succeeds even when trained with samples from other compositional systems. For example, a sample with a TaNbFeSb composition can be used as a training data point for a VNbFeSb; this is the case in Figure \ref{fig:transfer_learn}. The AL regression method can determine interactions between Nb, Fe and Sb from the TaNbFeSb and apply it to the VNbFeSb sample. 

Active learning methods have been previously applied to predict single scalar outputs from multivariate inputs \cite{Bishnoi2019, Tripathi2022}. Many materials engineering problems require simultaneous optimization of multiple parameters, whether it is a multinary composition or competing material properties like hardness and strength. Furthermore, many active learning studies have focused on synthesis methods that can be automated or performed in a high-throughput manner \cite{Mekki-Berrada2021}. The intersections of active learning, high throughput experimentation, and automation are currently being pursued by many self-driving laboratories around the world. The Ada laboratory at the University of British Columbia has demonstrated several successes in machine learning-driven material synthesis \cite{MacLeod2020}. Notably, the system found a global maxima of hole mobility in 35 sample runs. In another publication, authors demonstrated Ada for performing multi-objective optimization of thin films with competing objectives \cite{MacLeod2022}. 

Yet, there still exist a multitude of material synthesis methods that are not yet automated, or not easily converted into a high throughput method. In these cases, researchers are attempting to optimize multivariate objectives from sparse data. Active learning can potentially benefit these labor-intensive processes all the same. Human operators, acting alongside active learning algorithms, can reach optimized synthesis parameters much faster than a human operator alone. The high predictive performance of these models on sparse datasets enables researchers to use active learning with only a single sample needed.

This study builds upon these previous investigations but demonstrates active learning-based multi-output recommendations for targeted synthesis on a system that is not high throughput or robotically controlled but, instead, labor intensive and thus built on sparse and expensive measurements. The goal of this effort is not to replace a human operator with a robot but to augment human performance by quantifying the structure of the target space and identifying trends and insights that could be converted into rule-of-thumb to guide optimization when the empirical dataset is even sparser at the beginning of a synthesis campaign in a compositional space.    

\section{Conclusion}
Materials science is increasingly moving in the direction of large datasets, whether computationally generated or experimentally generated. These large datasets are well-equipped for deep learning algorithms such as Google's GNoME platform \cite{Merchant2023}. Concurrently, advances in robotic systems are pushing materials science towards automated synthesis and experimental discovery through self-driving laboratories. 

Yet, still, the vast majority of materials science research is not yet compatible with deep learning methods due to data sparsity and most laboratories do not have access to high throughput robotic systems. Active learning, together with transfer learning stand to fill this gap and enable higher productivity for these traditional laboratories. 

Although the AL approach was demonstrated for magnetron sputtering, it is extensible to other thin-film physical vapor deposition techniques such as other physical vapor deposition systems \cite{Kelly2000, Musila2005, Sarakinos2010, Liang2023}, chemical vapor deposition \cite{Yan2018, Zhou2022}, and plasma laser deposition \cite{Sloyan2009,Keller2015}. It may also be extensible to other manufacturing techniques that are high-throughput compatible such as additive manufacturing or continuous flow synthesis \cite{Dunlap2023}. This active learning workflow applies as long as the system has a finite number of tunable parameters, and those tunable parameters are strong predictors of sample composition. 

This method also has the potential to be incorporated as a `cold start' for other active learning workflows, such as those used in self-driving laboratories. The operation of autonomous workflows still requires an initial point of entry. Where to start regarding instrument process calibration is not always clear \textit{before} autonomous synthesis. The method detailed herein can be used to initialize synthesis workflows to achieve a targeted composition, or to efficiently explore compositional systems, by seeding with previously collected data even if the data is not completely in the target space, or is in a lower-dimensional target space. 

We strongly encourage researchers who are working on labor intensive synthesis where datasets are too sparse for many machine learning approaches to adopt active- and transfer-learning approaches in their day-to-day laboratory operations. Having humans work alongside active learning models can vastly improve synthesis efficiency, productivity, and throughput.

\section{Methods}
\subsection{Sample Preparation and Synthesis}
The samples synthesized in this study are all predicted to be half-Heusler (F$\bar 3$4m) thermoelectrics. Half-heuslers form at specific stoichiometries \cite{Anand2018}, specifically when the unit cell has a total of 18 valence electrons across all constituents. A three-element half-Heusler usually has equiatomic proportions. In a four-element half-Heusler, two elements occupy the first two Wyckoff sites and make up one-third of the atoms each. The other two atoms split occupancy on the third Wyckoff site and make up one-sixth of atoms each \cite{Zeier2016}.

The target compositions for the ternary system are equiatomic ratios, or A$_{0.3\bar3}$B$_{0.3\bar3}$C$_{0.3\bar3}$. For a quarternary alloy the target compositions are A$_{0.1\bar6}$B$_{0.1\bar6}$C$_{0.3\bar3}$D$_{0.3\bar3}$. For quinary compositions, the target is A$_{0.1\bar1}$B$_{0.1\bar1}$C$_{0.1\bar1}$D$_{0.3\bar3}$E$_{0.3\bar3}$. 

The sputtering system used was an AJA International Orion ATC system \cite{AJAInternational}. The system uses a dual turbopump and cryogenic pump to achieve ultra high vacuum. The chamber pressure before all depositions was $10^{-8}$ Torr. The system has six sputtering guns; two use a radio-frequency power source and four use a direct current power source. Elemental targets were sourced from Kurt J. Lesker \cite{KJLTargets} and were 2 inches in diameter. All non-magnetic targets had a thickness of 1/4 inch, while metallic targets were 1/12 inch thick. Films were deposited on undoped single crystal Si wafers with a $<100>$ orientation. (University Wafer). Wafers were nominally 380 $\mu$m thick and 3" diameter. All wafers were cleaned by acetone and electrostatically shocked by a radio frequency cathode at 100W before deposition to clean the surface of any contaminants.

A pre-sputter routine was used for every deposition.  The chamber is initially flooded with 30 mTorr of ultra high purity Argon and all cathode guns are powered on at a constant value; this initiates sputtering on each gun. After a few seconds, the pressure is decreased to 3 mTorr and the shutters on the sputtering guns are closed. The targets are allowed to sputter with the shutter closed for two minutes so that the sputtering rate reaches steady state. After two minutes the shutters are open and all active guns sputter for one hour. Film thicknesses vary depending on the sputtering rates of the individual elements and the material density. In general films have a thickness on the order of hundreds of nanometers.

\subsection{Sample Characterization}
Sample composition is analyzed using a JEOL JXA-8230 microprobe analyzer with wavelength dispersive spectroscopy \cite{JEOL}. A thin film correction term is calibrated and fit to the data for each sample \cite{Takakura1998}. Corrections are also made for peak overlap, depending on the composition system analyzed. Multiple WDS measurements are taken at different regions in the films to get an aggregate composition. The final composition reported is the average of all measurements for a single sample. Wavelength dispersive spectroscopy has shown to have accuracy to within $\pm 3$ atomic percent \cite{Abou-Ras2011}.

\subsection{Active Learning}
There are many applications where supervised learning may be helpful but access to labeled data is sparse and obtaining new labeled data is non-trivial. This is often the case in synthesis studies; manufacturing a new sample or measuring a sample's properties can be time or resources intensive. In many cases, it is both. Active learning regression models are trained on the currently labeled dataset, even if sparse, and the trained model is used to select the optimal input to be labeled. The `optimal' input can be the one that most improves the models predictive accuracy, lowers its uncertainty, or otherwise. Data labeling occurs in an interactive cycle where the model chooses samples that are most beneficial at each iteration. This contrasts strategies like uniform sampling or random sampling. Active learning enables machine learning models to achieve better performance with fewer labeled samples, by allowing the model to choose the data it learns from. In prior research, active learning has shown orders of magnitude reduction in the number of labeled samples to be generated to train a model with commensurate accuracy \cite{Lookman2019}. 

The primary elements of an active learning approach are a surrogate model and a query strategy. The surrogate model is the regression model used to make predictions. Surrogate models must provide some type of uncertainty or quality metrics so that sample optimality can be measured. The query strategy is the method used to determine the optimal next sample. There are different query strategies such as Uncertainty Based Sampling where the next sample to label is the one that the model is most uncertain about. Another is Querying by Committee where an ensemble of models is trained. The variance in predictions across the ensemble reflects the uncertainty in prediction and the optimal next sample reduces this uncertainty by the largest margin. A third example is Expected Model Change which uses model gradients to identify inputs with maximal expected gradient lengths. In this investigation we use Uncertainty Based Sampling with the Gaussian Process surrogate model and Querying By Committee with the Random Forest surrogate model.

A \textit{Gaussian Process} (GP) \cite{rasmussen2006gaussian} is a non-parametric model that calculates probability densities over the space of possible regression functions, offering a probabilistic model. Unlike a Gaussian distribution, which is characterized by a mean and covariance, a Gaussian process is defined through a normal distribution over mean and covariance functions, denoted as $Y \sim GP(m(X),k(X,X'))$. In this notation, $m(X)$ and $k(X,X')$ represent the mean and covariance functions. GP models can capture various complex relationships while providing credibility intervals for their predictions. Due to these advantages, they are considered the default proxy model in active learning and we employ them for Uncertainty Based Sampling.

In this investigation, the Gaussian process regression model was implemented using the sklearn library. Several different kernels were tested, including Matern, Rational Quadratic, and radial basis functions, as well as combinations of these three kernels. All kernels incorporated a homoscedastic white noise kernel. Optimal performance was achieved with a combination of a Matern and Rational Quadratic kernel, along with a white noise kernel. During each training step, the GPR was queried on all remaining untrained samples. As GP Regressors can be queried directly for uncertainty in predictions, the sample with the highest prediction uncertainty was selected as the next input for the model.

A Random Forest \cite{breiman2001random} is an ensemble model that employs a set of trained decorrelated Classification And Regression Trees (CARTs), achieved through Bootstrapping and Feature Bagging. This decorrelation ensures that the Random Forest has lower variance than the individual tree models while maintaining low bias, thus addressing the bias-variance tradeoff. The final prediction of the Random Forest is an average over the predictions of the trees in the ensemble. In the context of active learning, the variance between the predictions of the trained tree models in the Random Forest is taken as a measure of uncertainty in a Querying By Committee policy.

The random forest model used in this study was implemented using the sklearn library. The employed model consisted of 10 random forests, each containing 250 estimators. Larger models with up to 50 random forests and 500 estimators were tested; increasing the model size beyond 250 estimators and 10 random forests did not significantly improve predictive accuracy but substantially increased training time. For active learning, samples were selected to teach the model based on committee voting. All 10 tree models were queried for predictions on the remaining untrained samples. The sample exhibiting the highest variance in prediction was chosen as the next teaching input for the model.

A Neural Network (NN) model was also used, due to the popularity and widespread application of NN-based approaches across diverse disciplines. The neural network model performed worse than all other models and the human operator, both in terms of prediction accuracy and the number of samples required to achieve a given accuracy. This is indicative of the observation that Neural Network models, bereft of any additional Inductive Biases, need a higher number of samples to match the performance of classical ML approaches like Random Forest based models \cite{grinsztajn2022tree, mishra2021uncertainty, caruana2006empirical, rodriguez2015machine}. 

A \textit{neural network} model processes data through multiple interconnected layers that each perform mathematical operations on the input data to elucidate features and trends. Several different model architectures and hyperparameters were attempted. Fully connected or densely connected architectures, with either two or three hidden layers, performed very poorly. The best fully dense network attempted had a minimum MAE of 14\% after training on all samples in the dataset. 

The best-performing architecture (out of the ones tried) was an Attention-based architecture with three hidden layers. The rationale behind using an attention-based network is that for a given target system some samples are more important for training than others. For example, when manufactured an ABCD alloy, training on other ABCD alloys will yield better performance than training on CDEF alloys.  The first layer, a fully connected layer, utilized a Rectified Linear Unit (ReLU) activation function. Following this, the attention layer implemented a softmax function to compute the attention weights. A subsequent fully connected layer also used a ReLU activation function. Model weights were optimized via backpropagation and a custom loss function, which was based on the mean absolute difference between target and predicted values. The choice of an attention-based network was influenced by the input data's characteristics, as many sample vectors contained a substantial number of zeros. Attention-based neural networks allocate soft weights to input features according to their significance. In this dataset, specific parameters greatly affected the final composition due to their exceptionally high or low sputtering rates. Moreover, low-dimensional samples, such as ternary samples, contained numerous zero values that needed to be disregarded. Active learning with the neural network model was facilitated by committee voting. Five neural network models were trained on identical input data. After each training step, the five models made predictions on all samples in the untrained dataset. The sample with the highest variance in predicted value was selected as the next teaching input for the model.

\subsection{Transfer Learning}

In the active learning approach described above, new training samples are selected from anywhere within the composition space. This implies that when training a model to make predictions for 5-element samples, active learning models can suggest samples of lower dimensions for testing. However, these models will only make suggestions within the target composition space. It is essential to consider that information from overlapping composition spaces can be utilized to train the model. For instance, a manufacturer working with an ABC alloy may find useful correlations in a BCD alloy. Although the two composition systems share only two elements (B and C), if the model can determine the relationship between B and C in the ABC system, it can apply this information to make predictions for the BCD system.

Moreover, human operators commonly adopt an approach of working towards progressively more complex systems due to the challenging nature of tuning sputtering parameters. If an operator aims to create an ABCDE alloy, they might initially develop an ABC alloy. After fine-tuning the parameters for the ABC alloy, they proceed to create an ABCD alloy. Following further adjustments, they will attempt the ABCDE alloy. This method reduces complexity by introducing only one new element for tuning at a time. Attempting to create an ABCDE alloy without prior observations of sputtering element interactions can be disastrous.

Many laboratories have previously manufactured and measured samples from past experiments. It would be beneficial to leverage these samples to help initialize tuning for a new target system, even if the old samples do not share all of the same elements. To address this, a transfer learning approach was adopted. Initially, the active learning models are trained solely on ternary samples that share at least one element with the target system. Once all of the current ternary samples have been taught to the model, quarternary samples are added. Again, these quaternary samples must share at least one element with the target system. After the ternary samples have been integrated into the model, the conventional active learning approach is utilized. The model is queried within the target ABCDE system to identify the best training sample to reduce model uncertainty.

In the context of transfer learning, we employ a concept known as \textit{dummy dimensions}. This approach involves training an algorithm on all dimensions of a problem space, which, in this case, comprises six dimensions due to the six elements within the system. However, the input data typically contains only a few non-zero elements.

For a specific composition $y$, the algorithm receives a vector of length six. In the case of a ternary sample, only three out of the six entries contain non-zero values, and for a quarternary sample, four entries are non-zero, and so forth. The position of each element in the vector $y$ is preserved. For instance, the atomic percentage of Nb (Niobium) is always represented as the first entry in $y$. If there is no Nb in the sample, then the first entry is set to zero. Similarly, the second entry always corresponds to Titanium, and so on.

Initially, the model is trained using samples that have only three non-zero values for power, angle, and atomic percentage. After the model is proficient with these samples, it proceeds to train on samples with four non-zero values for power, angle, and atomic percentage. This process continues for samples with five, and so on.

Several different transfer learning models were trained, as shown in Figure \ref{fig:transfer_learn}. Three models were trained on only one compositional subsystem (ternary only, quarternary only, and quinary only). A random sample was chosen from the manifold as the initial training point then further samples were selected based on maximum uncertainty. 

Next, models were created that were trained on the entirety of the ternary manifold (15 samples) and used to make predictions on the quarternary or quinary dataset. Again, one initial sample was chosen from the target manifold (quarternary or quinary) and further samples were selected using maximum uncertainty sampling. 

Finally, a model was trained on the entirety of both the ternary and quarternary manifolds (30 samples) and predictions were made on the quinary dataset. An random quinary sample was chosen as the initial training point and then sampling proceeded again using maximum uncertainty.

All models shown in the paper were re-run 10 times with a different random sampling of the manifold for initial training points (5 for the full model, 1 for the transfer learning models). The MAE reported in Figures \ref{fig:all_models_total_errors} and \ref{fig:transfer_learn} are the average errors across all runs. The error bars on the MAE represent the standard deviation in MAE across all 10 training runs. 

\section*{Acknowledgements}
This research was supported by the U.S. Department of Energy, Office of Energy Efficiency and Renewable Energy (EERE), specifically the Advanced Materials \& Manufacturing Technologies Office (AMMTO), under contract DE-AC02-76SF00515. Aashwin Mishra was partially supported by the SLAC ML Initiative.

\section*{Author Contributions}
NSJ: conceptualization, material synthesis, material characterization, data collection,  algorithm development, writing -- all drafts, visualization, review, editing.
AAM: conceptualization, algorithm development, writing -- all drafts, visualization, review, editing.
AM: supervision, project administration, writing -- review, editing.

\section*{Data Availability}
All data and associated code to reproduce this data is included as supplemental information.

\bibliographystyle{ieeetr}
\bibliography{bib.bib}

\begin{thebibliography}{10}

\bibitem{Kim2019}
Y.~S. Kim, H.~J. Park, S.~C. Mun, E.~Jumaev, S.~H. Hong, G.~Song, J.~T. Kim,
  Y.~K. Park, K.~S. Kim, S.~I. Jeong, and et~al., ``Investigation of structure
  and mechanical properties of \text{TiZrHfNiCuCo} high entropy alloy thin
  films synthesized by magnetron sputtering,'' {\em Journal of Alloys and
  Compounds}, vol.~797, p.~834–841, 2019.

\bibitem{Rar2004}
A.~Rar, J.~J. Frafjord, J.~D. Fowlkes, E.~D. Specht, P.~D. Rack, M.~L.
  Santella, H.~Bei, E.~P. George, , and G.~M. Pharr, ``\text{PVD} synthesis and
  high-throughput property characterization of \text{Ni–Fe–Cr} alloy
  libraries,'' {\em Measurement Science and Technology}, vol.~16, p.~834–841,
  2019.

\bibitem{Wang2022}
K.~Wang, K.~Nishio, K.~Horiba, M.~Kitamura, K.~Edamura, D.~Imazeki,
  R.~Nakayama, R.~Shimizu, H.~Kumigashira, and T.~Hitosugi, ``Synthesis of
  high-entropy layered oxide epitaxial thin films:
  \text{LiCr1/6Mn1/6Fe1/6Co1/6Ni1/6Cu1/6O2},'' {\em Crystal Growth and Design},
  vol.~22, p.~1116–1122, January 2022.

\bibitem{Anand2018}
S.~Anand, K.~Xia, V.~I. Hegde, U.~Aydemir, V.~Kocevski, T.~Zhu, C.~Wolverton,
  and G.~J. Snyder, ``A valence balanced rule for discovery of 18-electron
  half-heuslers with defects,'' {\em Energy and Environmental Science},
  vol.~11, 2018.

\bibitem{Zarnetta2010}
R.~Zarnetta, ``Identification of quarternary shape memory alloys with near-zero
  thermal hysteresis and unprecedented functional stability,'' {\em Advanced
  Functional Materials}, vol.~20, pp.~1917--1923, 2010.

\bibitem{AlHasan2020}
N.~M.~A. Hasan, H.~Hou, S.~Sarkar, S.~Thienhaus, A.~Mehta, A.~Ludwig, and
  I.~Takeuchi, ``Combinatorial synthesis and high-throughput characterization
  of microstructure and phase transformation in \text{Ni-Ti-Cu-V} quarternary
  thin-film library,'' {\em Engineering}, vol.~6, no.~6, pp.~637--643, 2020.

\bibitem{Li2022}
S.-Y. Li, T.~X. Nguyen, Y.-H. Su, C.-C. Lin, Y.-J. Huang, Y.-H. Shen, C.-P.
  Liu, J.-J. Ruan, K.-S. Chang, and J.-M. Ting, ``Sputter-deposited high
  entropy alloy thin film electrocatalyst for enhanced oxygen evolution
  reaction performance,'' {\em Small}, August 2022.

\bibitem{Jung2022}
S.-G. Jung, Y.~Han, J.~H. Kim, R.~Hidayati, J.-S. Rhyee, J.~M. Lee, W.~N.~K.
  nd~Woo Seok~Choi, H.-R. Jeon, J.~Suk, and T.~Park, ``High critical current
  density and high-tolerance superconductivity in high-entropy alloy thin
  films,'' {\em Nature Communications}, vol.~13, no.~3373, 2022.

\bibitem{Yan2018}
C.~Han, J.~Zhi, Z.~Zeng, Y.~Wang, B.~Zhou, J.~Gao, Y.~Wu, Z.~He, X.~Wang, and
  S.~Yu, ``Synthesis and characterization of nano-polycrystal diamonds on
  refractory high entropy alloys by chemical vapour deposition,'' {\em
  Materials Chemistry and Physics}, vol.~210, pp.~12--19, May 2018.

\bibitem{Zhou2022}
B.~Zhou, Y.~Wang, C.~Xue, C.~Han, H.~Hei, Y.~Xue, Z.~Liu, Y.~Wu, Y.~Ma, J.~Gao,
  and S.~Yu, ``Chemical vapor deposition diamond nucleation and initial growth
  on \text{TiZrHfNb} and \text{TiZrHfNbTa} high entropy alloys,'' {\em
  Materials Letters}, vol.~309, February 2022.

\bibitem{Liang2023}
A.~Liang, D.~C. Goodelman, A.~M. Hodge, D.~Farkas, and P.~S. Branicio,
  ``\text{CoFeNiTi}$_x$ and \text{CrFeNiTi}$_x$ high entropy alloy thin films
  microstructure formation,'' {\em Acta Materialia}, vol.~257, September 2023.

\bibitem{Han2023}
C.~Han, J.~Zhi, Z.~Zeng, Y.~Wang, B.~Zhou, J.~Gao, Y.~Wu, Z.~He, X.~Wang, and
  S.~Yu, ``Synthesis and characterization of nano-polycrystal diamonds on
  refractory high entropy alloys by chemical vapour deposition,'' {\em Applied
  Surface Science}, vol.~623, June 2023.

\bibitem{Oses2020}
C.~Oses, C.~Toher, and S.~Curtarolo, ``High-entropy ceramics,'' {\em Nature
  Reviews Materials}, vol.~5, p.~295–309, 2020.

\bibitem{Gregoire2023}
J.~M. Gregoire, L.~Zhou, and J.~A. Haber, ``Combinatorial synthesis for
  ai-driven materials discovery,'' {\em Nature Synthesis}, vol.~2,
  p.~493–504, 2023.

\bibitem{Bunn2016}
J.~K. Bunn, R.~Z. Voepel, Z.~Wang, E.~P. Gatzke, J.~A. Lauterbach, , and J.~R.
  Hattrick-Simpers, ``Development of an optimization procedure for
  magnetron-sputtered thin films to facilitate combinatorial materials
  research,'' {\em Industrial \& Engineering Chemical Research}, vol.~55,
  p.~1236–1242, 2016.

\bibitem{Xia2020}
A.~Xia, A.~Togni, S.~Hirn, G.~Bolelli, L.~Lusvarghi, and R.~Franz,
  ``Angular-dependent deposition of \text{Mo}\text{Nb}\text{Ta}\text{V}\text{W}
  \text{HEA} thin films by three different physical vapor deposition methods,''
  {\em Surface and Coatings Technology}, vol.~385, March 2020.

\bibitem{Alami2006}
J.~Alami, P.~Eklund, J.~Emmerlich, O.~Wilhelmsson, U.~Jansson, H.~Högberg,
  L.~Hultman, and U.~Helmersson, ``High-power impulse magnetron sputtering of
  \text{Ti–Si–C} thin films from a \text{Ti}$_3$\text{SiC}$_2$ compound
  target,'' {\em Thin Solid Films}, vol.~515, p.~1731–1736, 2006.

\bibitem{Deki1997}
S.~Deki, Y.~Aoi, Y.~Asaoka, A.~Kajinami, and M.~Mizuhata, ``Monitoring the
  growth of titanium oxide thin films by the liquid-phase deposition method
  with a quartz crystal microbalance,'' {\em Journal of Materials Chemistry},
  vol.~7, pp.~733--736, 1997.

\bibitem{Gongora2020}
A.~E. Gongora, B.~Xu, W.~Perry, C.~Okoye, P.~Riley, K.~G. Reyes, E.~F. Morgan,
  and K.~A. Brown, ``A {B}ayesian experimental autonomous researcher for
  mechanical design,'' {\em Science Advances}, vol.~6, April 2020.

\bibitem{Xue2016}
D.~Xue, P.~V. Balachandran, J.~Hogden, J.~Theiler, D.~Xue, and T.~Lookman,
  ``Accelerated search for materials with targeted properties by adaptive
  design,'' {\em Nature Communications}, vol.~7, no.~11241, 2016.

\bibitem{Kusne2020}
A.~G. Kusne, H.~Yu, C.~Wu, H.~Zhang, J.~Hattrick-Simpers, B.~DeCost, S.~Sarker,
  C.~Oses, C.~Toher, S.~Curtarolo, A.~V. Davydov, R.~Agarwal, L.~A. Bendersky,
  M.~Li, A.~Mehta, and I.~Takeuchi, ``On-the-fly closed-loop materials
  discovery via bayesian active learning,'' {\em Nature Communications},
  vol.~11, 2020.

\bibitem{Nikolaev2016}
P.~Nikolaev, D.~Hooper, F.~Webber, K.~D. Rahul~Rao, M.~Krein, J.~Poleski,
  R.~Barto, and B.~Maruyama, ``Autonomy in materials research: a case study in
  carbon nanotube growth,'' {\em npj Computational Materials}, vol.~2,
  no.~16031, 2016.

\bibitem{Ament2023}
S.~Ament, M.~Amsler, D.~R. Sutherland, M.-C. Chang, D.~Guevarra, A.~B.
  Connolly, J.~M. Gregoire, M.~O. Thompson, C.~P. Gomes, and R.~B.~V. Dover,
  ``Autonomous materials synthesis via hierarchical active learning of
  nonequilibrium phase diagrams,'' {\em Science Advances}, vol.~7, Dec 2021.

\bibitem{grinsztajn2022tree}
L.~Grinsztajn, E.~Oyallon, and G.~Varoquaux, ``Why do tree-based models still
  outperform deep learning on typical tabular data?,'' {\em Advances in Neural
  Information Processing Systems}, vol.~35, pp.~507--520, 2022.

\bibitem{caruana2006empirical}
R.~Caruana and A.~Niculescu-Mizil, ``An empirical comparison of supervised
  learning algorithms,'' in {\em Proceedings of the 23rd international
  conference on Machine learning}, pp.~161--168, 2006.

\bibitem{rasmussen2006gaussian}
C.~E. Rasmussen and C.~K. Williams, {\em Gaussian processes for machine
  learning}, vol.~1.
\newblock Springer, 2006.

\bibitem{kraskov2004estimating}
A.~Kraskov, H.~St{\"o}gbauer, and P.~Grassberger, ``Estimating mutual
  information,'' {\em Physical review E}, vol.~69, no.~6, p.~066138, 2004.

\bibitem{arrieta2020explainable}
A.~B. Arrieta, N.~D{\'\i}az-Rodr{\'\i}guez, J.~Del~Ser, A.~Bennetot, S.~Tabik,
  A.~Barbado, S.~Garc{\'\i}a, S.~Gil-L{\'o}pez, D.~Molina, R.~Benjamins, {\em
  et~al.}, ``Explainable artificial intelligence (xai): Concepts, taxonomies,
  opportunities and challenges toward responsible ai,'' {\em Information
  fusion}, vol.~58, pp.~82--115, 2020.

\bibitem{adhikari2019leafage}
A.~Adhikari, D.~M. Tax, R.~Satta, and M.~Faeth, ``Leafage: Example-based and
  feature importance-based explanations for black-box ml models,'' in {\em 2019
  IEEE international conference on fuzzy systems (FUZZ-IEEE)}, pp.~1--7, IEEE,
  2019.

\bibitem{davis2020measure}
B.~Davis, M.~Glenski, W.~Sealy, and D.~Arendt, ``Measure utility, gain trust:
  practical advice for xai researchers,'' in {\em 2020 IEEE workshop on trust
  and expertise in visual analytics (TREX)}, pp.~1--8, IEEE, 2020.

\bibitem{Bishnoi2019}
S.~Bishnoi, S.~Singha, R.~Ravindera, M.~Bauchyb, N.~N. Gosvamic, H.~Kodamanad,
  and N.~A. Krishnan, ``Predicting young's modulus of oxide glasses with sparse
  datasets using machine learning,'' {\em Journal of Non-Crystalline Solids},
  vol.~524, no.~119643, 2019.

\bibitem{Tripathi2022}
B.~M. Tripathi, A.~Sinha, and T.~Mahata, ``Machine learning guided study of
  composition-coefficient of thermal expansion relationship in oxide glasses
  using a sparse dataset,'' {\em Materials Today: Proceedings}, vol.~67,
  pp.~326--329, 2022.

\bibitem{Mekki-Berrada2021}
F.~Mekki-Berrada, Z.~Ren, T.~Huang, W.~K. Wong, F.~Zheng, J.~Xie, I.~P.~S.
  Tian, S.~Jayavelu, Z.~Mahfoud, D.~Bash, K.~Hippalgaonkar, S.~Khan,
  T.~Buonassisi, Q.~Li, and X.~Wang, ``Two-step machine learning enables
  optimized nanoparticle synthesis,'' {\em npj Computational Materials},
  vol.~7, no.~55, 2021.

\bibitem{MacLeod2020}
B.~P. MacLeod, F.~G.~L. Parlane, T.~D. Morrisey, F.~Hase, L.~M. Roch, K.~E.
  Dettelbach, R.~Moreira, L.~P.~E. Yunker, M.~B. Rooney, J.~R. Deeth, V.~Lai,
  G.~J. Ng, H.~Situ, R.~H. Zhang, M.~S. Elliot, T.~H. Haley, D.~J. Dvorak,
  A.~Aspuru-Guzik, J.~E. Hein, and C.~P. Berlinguette, ``Self-driving
  laboratory for accelerated discovery of thin-film materials,'' {\em Science
  Advances}, vol.~6, May 2020.

\bibitem{MacLeod2022}
B.~P. MacLeod, F.~G.~L. Parlane, C.~C. Rupnow, K.~E. Dettelbach, M.~S. Elliott,
  T.~D. Morrissey, T.~H. Haley, O.~Proskurin, M.~B. Rooney, N.~Taherimakhsousi,
  D.~J. Dvorak, H.~N. Chiu, C.~E.~B. Waizenegger, K.~Ocean, M.~Mokhtari, and
  C.~P. Berlinguette, ``A self-driving laboratory advances the pareto front for
  material properties,'' {\em Nature Communications}, vol.~13, no.~995, 2022.

\bibitem{Merchant2023}
A.~Merchant, S.~Batzner, S.~S. Schoenholz, M.~Aykol, G.~Cheon, and E.~D. Cubuk,
  ``Scaling deep learning for materials discovery,'' {\em Nature}, vol.~624,
  p.~80–85, 2023.

\bibitem{Kelly2000}
P.~Kelly and R.~Arnell, ``Magnetron sputtering: a review of recent developments
  and applications,'' {\em Vacuum}, vol.~56, pp.~159--172, 2000.

\bibitem{Musila2005}
J.~Musila, P.~Barocha, J.~Vlc\u{e}ka, K.~Namc, and J.~Hanc, ``Reactive
  magnetron sputtering of thin films: present status and trends,'' {\em Thin
  Solid Films}, vol.~475, p.~208 – 218, 2005.

\bibitem{Sarakinos2010}
K.~Sarakinos, J.~Alami, and S.~Konstantinidis, ``High power pulsed magnetron
  sputtering: A review on scientific and engineering state of the art,'' {\em
  Surface \& Coatings Technology}, vol.~204, p.~1661–1684, 2010.

\bibitem{Sloyan2009}
K.~A. Sloyan, T.~C. May-Smith, R.~W. Eason, and J.~G. Lunney, ``The effect of
  relative plasma plume delay on the properties of complex oxide films grown by
  multi-laser, multi-target combinatorial pulsed laser deposition,'' {\em
  Applied Surface Science}, vol.~255, pp.~9066--9070, August 2009.

\bibitem{Keller2015}
D.~A. Keller, A.~Ginsburg, H.-N. Barad, K.~Shimanovich, Y.~Bouhadana,
  E.~Rosh-Hodesh, I.~Takeuchi, H.~Aviv, Y.~R. Tischler, A.~Y. Anderson†, and
  A.~Zaban, ``Utilizing pulsed laser deposition lateral inhomogeneity as a tool
  in combinatorial material science,'' {\em ACS Combinatorial Science},
  vol.~17, p.~209–21, March 2015.

\bibitem{Dunlap2023}
J.~H. Dunlap, J.~G. Ethier, A.~A. Putnam-Neeb, S.~Iyer, S.-X.~L. Luo, H.~Feng,
  J.~A.~G. Torres, A.~G. Doyle, T.~M. Swager, R.~A. Vaia, P.~Mirau, C.~A.
  Crouse, and L.~A. Baldwin, ``Continuous flow synthesis of pyridinium salts
  accelerated by multi-objective bayesian optimization with active learning,''
  {\em Chemical Science}, vol.~14, pp.~8061--8069, 2023.

\bibitem{Zeier2016}
W.~G. Zeier, J.~Schmitt, G.~Hautier, U.~Aydemir, Z.~M. Gibbs, C.~Felser, and
  G.~J. Snyder, ``Engineering half-heusler thermoelectric materials using zintl
  chemistry,'' {\em Nature Materials}, vol.~1, May 2016.

\bibitem{AJAInternational}
A.~International, ``{ATC} {O}rion {M}agnetron {S}puttering system.''
  https://www.ajaint.com/atc-orion-series-sputtering-systems.html, 2023.

\bibitem{KJLTargets}
\text{Kurt J. Lesker}, ``Sputtering {T}argets.''
  www.lesker.com/materials-division.cfm/section-sputtering-targets, 2023.

\bibitem{JEOL}
\text{JEOL Ltd.}, ``\text{JEOL JXA-8230}.''
  https://www.jeol.com/products/scientific/epma/, 2023.

\bibitem{Takakura1998}
M.~Takakura, H.~Takahashi, and T.~Okumura, {\em Thin-Film Analysis with
  Electron Probe X-ray MicroAnalyzer}, vol.~33E.
\newblock 1998.

\bibitem{Abou-Ras2011}
D.~Abou-Ras, R.~Caballero, C.-H. Fischer, C.~Kaufmann, I.~Lauermann, R.~Mainz,
  H.~Mönig, A.~Schöpke, C.~Stephan, C.~Streeck, S.~Schorr, A.~Eicke,
  M.~D\"obeli, B.~Gade, J.~Hinrichs, T.~Nunney, H.~Dijkstra, V.~Hoffmann,
  D.~Klemm, V.~Efimova, A.~Bergmaier, G.~Dollinger, T.~Wirth, W.~Unger,
  A.~Rockett, A.~Perez-Rodriguez, J.~Alvarez-Garcia, V.~Izquierdo-Roca,
  T.~Schmid, P.-P. Choi, M.~Müller, F.~Bertram, J.~Christen, H.~Khatri,
  R.~Collins, S.~Marsillac, and I.~Kötschau, ``Comprehensive comparison of
  various techniques for the analysis of elemental distributions in thin
  films,'' {\em Microscopy and Microanalysis}, vol.~17, p.~728–751, 2011.

\bibitem{Lookman2019}
T.~Lookman, P.~V. Balachandran, D.~Xue, and R.~Yuan, ``Active learning in
  materials science with emphasis on adaptive sampling using uncertainties for
  targeted design,'' {\em npj Computational Materials}, February 2019.

\bibitem{breiman2001random}
L.~Breiman, ``Random forests,'' {\em Machine learning}, vol.~45, pp.~5--32,
  2001.

\bibitem{mishra2021uncertainty}
A.~A. Mishra, A.~Edelen, A.~Hanuka, and C.~Mayes, ``Uncertainty quantification
  for deep learning in particle accelerator applications,'' {\em Physical
  Review Accelerators and Beams}, vol.~24, no.~11, p.~114601, 2021.

\bibitem{rodriguez2015machine}
V.~Rodriguez-Galiano, M.~Sanchez-Castillo, M.~Chica-Olmo, and M.~Chica-Rivas,
  ``Machine learning predictive models for mineral prospectivity: An evaluation
  of neural networks, random forest, regression trees and support vector
  machines,'' {\em Ore Geology Reviews}, vol.~71, pp.~804--818, 2015.

\end{thebibliography}

\end{document}


\title{Supplemental Information}

\begin{figure}
    \includegraphics[width=1\linewidth]{all_norms_joint_probability_06262023.png}
    \label{The joint normal probability distribution of all gun powers with each other. Each subplot shows the joint probability distribution of one elements sputtering power with another. The histogram bars on the top and right side of the plots show the number of measurements made at that power.}
    \label{joint_probability}
\end{figure}

\begin{figure}
    \includegraphics[width=1\linewidth]{lsq_regression_error.png}
    \caption{Performance of linear least-squares regression on the composition manifold.}
    \label{lsq_regression_error}
\end{figure}